# Gaussian fluctuations in an ideal bose-gas – a simple model

A Petrova[1,2], O Nedopekin[2], D Tayurskii[2] and Q A Wang[1]

[1]LUNAM Universite, ISMANS, Laboratoire de Physique Statistique et Systeme Complexes, 44, Avenue F.A. Bartholdi, Le Mans 72000, France

[2]Institute of Physics, Kazan Federal University, Kazan 420008, Russia

E-mail: npetrova131@gmail.com

**Abstract**. Based on the canonical ensemble, we suggested the simple scheme for taking into account Gaussian fluctuations in a finite system of ideal boson gas. Within framework of scheme we investigated the influence of fluctuations on the particle distribution in Bose -gas for two cases - with taking into account the number of particles in the ground state and without this assumption. The temperature and fluctuation parameter dependences of the modified Bose-Einstein distribution have been determined. Also the dependence of the condensation temperature on the fluctuation distribution parameter has been obtained.

## 1. Introduction

The experimental possibilities of detecting few and even single atoms confined in magnetic or optical traps with high precision open new perspectives to study the correlations and fluctuations in atomic and molecular gases at the mesoscopic level. Studying fluctuations in statistical systems is important from several points of view. For instance, the stability of a system and its way of reaching the equilibrium state depend drastically on the value of fluctuations [1].

The fluctuations in a Bose-Einstein condensate have been studied in a systematic way in the case of the ideal Bose gas in a box [2-4], and, more recently, in the presence of a harmonic trap [5-7]. The uniform ideal gas is an unstable system, with anomalously strong particle fluctuations. Switching on atomic interactions stabilizes the system, making the fluctuations normal. But further strengthening interactions leads to the condensate depletion. Contrary to the uniform case, the trapped ideal Bose gas is stable, having normal fluctuations. Atomic interactions play only the destructive role, depleting the condensate [8].

The aim of present research is the studying of the overlapping effects coming from Gaussian noise and quantum statistics of the ideal Bose-gas. We investigate the modification of Bose-Einstein distribution due to the thermal fluctuations obeying to the normal distribution and the changes of the dependence of Bose-Einstein condensation temperature on the fluctuation distribution parameter.

In the vicinity of critical points, there always exists a region, where the system experiences strong fluctuations [1]. In finite systems, the phase transition is of crossover type. Around the point of crossover, there can arise strong fluctuations, similarly to those appearing in the vicinity of the critical points. Then the region around a crossover point, where fluctuations are strong, can be called crossover region. Bose-Einstein condensation of trapped atoms is always a crossover. Considering phase transitions under varying temperature, one usually defines the critical region by means of the Ginzburg temperature $T_G$, such that in the critical region $T_G < T < Tc$ fluctuations are strong, while essentially below $T_G$, they are suppressed [9]. In general, one may distinguish the longitudinal and transverse fluctuations of an order parameter. The transverse, or directional, fluctuations often survive much longer, when lowering temperature below $Tc$, than the longitudinal, or size, fluctuations. Therefore the actual critical region is characterized rather by the transverse then longitudinal fluctuations [10].

In Bose-Einstein condensates, one may also distinguish two types of fluctuations. The size fluctuations of the order parameter correspond to density fluctuations. The transverse fluctuations are usually associated with phase fluctuations. In the vicinity of the condensation temperature $Tc$, both these fluctuations are strong. Lowering temperature below $Tc$ first diminishes the density fluctuations, while the phase fluctuations remain yet strong. Lowering temperature further, one comes to a point $T_f$, where phase fluctuations are getting suppressed as well. Thus, the crossover region is defined by the temperature interval $T_f < T < Tc$. Density fluctuations always excite current fluctuations. However, there can exist current fluctuations those do not initiate density fluctuations. This explains why the crossover region is mainly defined by the phase (current) fluctuations, but not by the density fluctuations.

The development of physics at the mesoscopic level and its applications (primarily in the field of nanotechnology) is associated with an increase of interest in learning more and more chaotic (including low-dimensional) systems. For such systems, even in thermal equilibrium or close to it the fluctuations of physical quantities are significant. But till now the theoretical description of the fluctuations influence cannot be considered as satisfactory. We should note here that fluctuations cannot be studied thermodynamically at extremely low temperatures or in the case of very rapid changes of their magnitude. Also at sufficient low temperatures pure quantum fluctuations begin to play an important role.

## 2. The study of the noise effect on the Bose-Einstein distribution

Statistics of weakly interacting Bose gas attracted much attention within recent years due to the experimental realization of Bose-Einstein condensation (BEC) of trapped gases. Statistical physics of Bose-Einstein condensation (BEC) is usually considered by using the grand canonical ensemble (thermodynamic coordinates $\mu$, $V$, $T$) in the thermodynamic limit, i.e., for a macroscopic system with the infinite number of particles and volume, but the finite concentration. However, the trapped BEC experiments deal with the finite number of particles ($N \sim 10^2$-$10^6$), which makes the system mesoscopic. This is the first reason to develop a microscopic theory of BEC in mesoscopic systems by means of the canonical ensemble (thermodynamic coordinates $N$, $V$, $T$) [11].

The grand-canonical description is not appropriate for the analysis of BEC fluctuations, even in the case of mesoscopic ideal Bose gas [12]. For example, the widely used grand-canonical approach yields the grand-canonical catastrophe. The restricted ensemble approach, which fixes the amplitude and the phase of the condensate wave function, is unable to analyze the fluctuation problem at all. To study fluctuations, one should fix only external macroscopic and global topological parameters of the system, such as temperature, pressure, number of particles, superfluid flow pattern, boundary conditions, etc. Such theory is more complex than it was believed commonly, because the result must clearly lead to the dependence on a finite number of particles in the system and its size.

In present work we are interested in the effects of noise on the Bose-Einstein distribution. To study these effects we propose to use the convolution of Bose-Einstein and Gaussian distributions. The main problem consists in the normalization of the convolution of quantum Bose-Einstein statistics and Gaussian function of noise. The standard expression for probability of state with energy $\varepsilon$ is

$$n_\alpha(\varepsilon) = \frac{1}{Z} \int_0^\infty n(x) f_\alpha(x - \varepsilon) dx , \qquad (1)$$

where $Z$ is normalization coefficient that could be found from this condition:

$$\int_0^\infty d\varepsilon n(\varepsilon)g(\varepsilon)=1, \qquad (2)$$

$n(\varepsilon)$ stands for the quantum distribution, $g(\varepsilon)$ denotes the density of states, $f_\alpha(x-\varepsilon)=(2\pi\sigma)^{-1/2}\exp(-(x-\varepsilon)^2/2\sigma^2)$ is the usual Gaussian distribution.

This normalization (by means of Z) is straightforward for case without fluctuations. But when fluctuations are introduced we have to take into account the change of chemical potential due to the noise. In self-consistent calculations the normalization will be done by means of the new value of the chemical potential. It is important to note here that the average number of particles in the system is conserved such kind of calculations.

Now we illustrate the proposed procedure for the case of ideal boson-gas, having $N$ particles and confined to the box of volume $V$. As usually the density of states in 3D space can be written as:

$$\sum_{\vec{k}} \frac{1}{e^{\frac{\varepsilon(\vec{k})-\mu}{\theta}}-1} \xrightarrow{3D} \frac{V}{(2\pi)^3}\int_0^\infty \frac{1}{e^{\frac{\varepsilon(\vec{k})-\mu}{\theta}}-1}d^3\vec{k} = \frac{V}{(2\pi)^3}\int_0^\infty d\vec{k}\vec{k}^2\int_\Omega d\Omega \frac{1}{e^{\frac{\varepsilon(\vec{k})-\mu}{\theta}}-1} =$$
$$= \frac{V}{(2\pi)^2}\left(\frac{2m}{\hbar^2}\right)^{3/2}\int_0^\infty \frac{\sqrt{\varepsilon}}{e^{\frac{\varepsilon(\vec{k})-\mu}{\theta}}-1}d\varepsilon, \qquad (3)$$

where $\varepsilon(\vec{k})=\frac{\hbar^2\vec{k}^2}{2m}$. Thus, density of states has the form:

$$g(\varepsilon)=\frac{V}{(2\pi)^2}\left(\frac{2m}{\hbar^2}\right)^{3/2}\sqrt{\varepsilon}=\frac{\sqrt{\varepsilon}}{\varepsilon_0^{3/2}}, \qquad (4)$$

where

$$\varepsilon_0=\left(\frac{4\pi^2}{V}\right)^{2/3}\times\frac{\hbar^2}{2m} \qquad (5)$$

and $m$ is the mass of free particle.

Using the general expression for Bose-Einstein distribution

$$n(\varepsilon)=\int_0^\infty \frac{g(\varepsilon)}{e^{\frac{\varepsilon-\mu}{kT}}-1}d\varepsilon \qquad (6)$$

one can write the convolution of the Gaussian noise and (6) for the total number of particles in the system in the following form:

$$N=\int_0^\infty\int_0^\infty \frac{f(x-\varepsilon)}{\exp(\frac{x-\mu}{kT})-1}dxg(\varepsilon)d\varepsilon, \qquad (7)$$

where

$$f(x-\varepsilon)=\frac{1}{\sigma\sqrt{\frac{\pi}{2}\left(1+erf\left[\frac{\varepsilon}{\sigma\sqrt{2}}\right]\right)}}\exp\left[-\frac{(x-\varepsilon)^2}{2\sigma^2}\right] \qquad (8)$$

represents the probability density of fluctuations with variance $\sigma^2$.

The next step is the finding of a new value of the chemical potential under assumption of the conservation of the total particle number (7). It is worth to note that all values of energy $x$, $\varepsilon$, fluctuation parameter $\sigma$, temperature $\theta = kT$ in expression (7) are measured in energetic units. Now we introduce new unit of energy $\varepsilon_0 = \left(4\pi^2/V\right)^{2/3} \times (\hbar^2/2m)$ and will measure all energy units (fluctuation parameter and temperature) by this quantity. The numerical calculations of the fluctuation effects were carried out by use the *Mathematica* software package.

The calculated particle distributions over energy at various temperatures are shown on Figure 1 for the ideal bose-gas with $N=100$. The analysis of the obtained data allows to make the following conclusions. As the system temperature decreases the probability to find the atoms in the ground state increases. Maximum of the particle's distribution shifts to lower energies with decrease the temperature. The Bose-Einstein distribution diffuses at high temperatures.

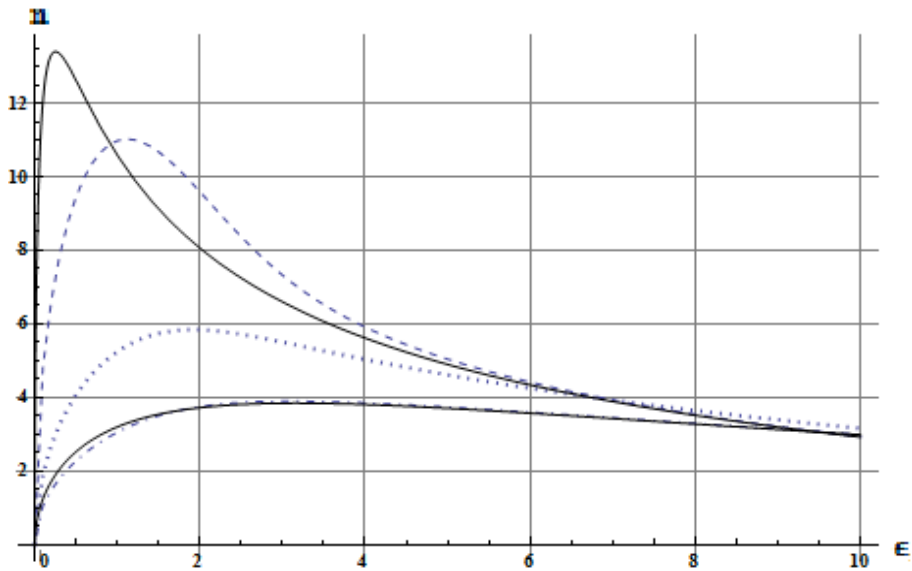

**Figure 1.** The particle number distributions over energy for different temperatures $T$. The upper solid line is the usual Bose-Einstein distribution at $T=14\varepsilon_0$, lower solid line is the usual Bose-Einstein distribution at $T=20\varepsilon_0$; dashed, dotted and chain lines show the distributions modified by fluctuations at temperatures $14\varepsilon_0, 17\varepsilon_0, 20\varepsilon_0$ respectively ($N=100$, $\sigma=1$).

In the figure 2 the calculated particle distributions over energy at various values of fluctuation parameter $\sigma$ are shown.

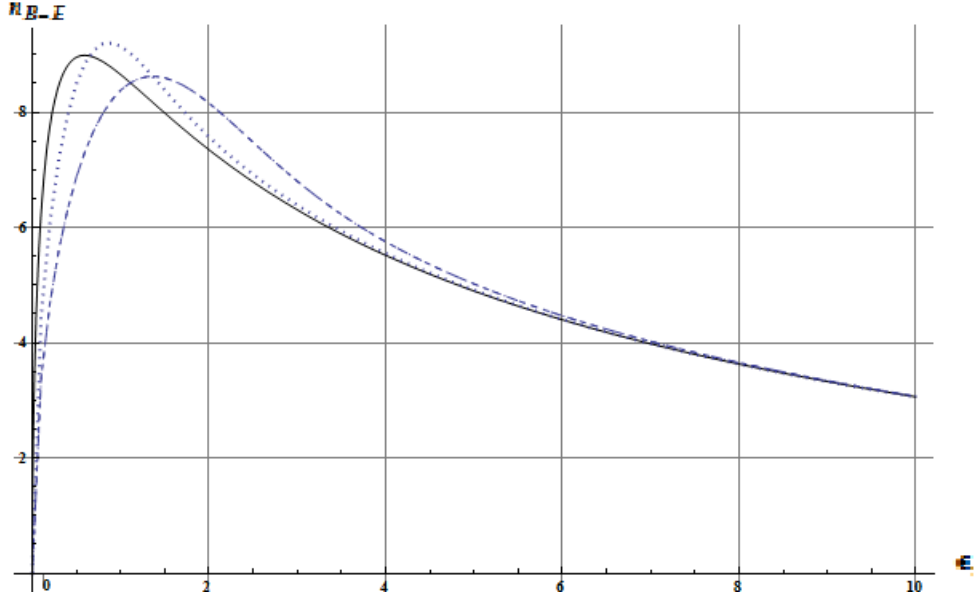

**Figure 2.** The modified by fluctuations particle distributions over energy for different values $\sigma$. The solid, dotted and chain lines of Bose-Einstein distributions at $\sigma = 0.1, 0.5, 1.0$, respectively ($N = 100$ and $T = 15\varepsilon_0$). For comparison the black color line shows the distribution of particles without fluctuations.

### 3. The study of the noise effect on the Bose-Einstein condensation

As noted before the chemical potential of the Bose-gas is always negative, and $\mu$ ($\theta = 0$) = 0 (the energy of the particles is measured from zero). This means that as temperature decreases the chemical potential increases (decreases by absolute value) and at a certain temperature $\theta_0$ it reaches zero value and remains equal to zero up to zero temperature. In thermodynamic limit at lowering the temperature in the range closed to $\mu = 0$ the number of particles in all states except the ground one is reduced drastically. This process is called Bose-condensation. To take into account the macroscopic number of particles in the ground state it is usual to separate the large term in the expression for the total number of particles in bose-gas [13]:

$$N = N_0(\theta) + \int_0^\infty \frac{g(\varepsilon)d\varepsilon}{\exp((\varepsilon-\mu)/\theta)-1} = N_0(\tau) + \frac{\gamma V M^{3/2}}{\sqrt{2}\pi^2\hbar^3}\zeta\left(\frac{3}{2}\right)\Gamma\left(\frac{3}{2}\right)\theta^{3/2}. \quad (9)$$

This expression is used in the standard approach for determining the number of particles in the Bose-Einstein condensate. Let's suppose that there are not fluctuations in the ground state, otherwise they destroy the condensate. In the case with fluctuations the total number of particles in the system is written as follows:

$$N = N_0 + \int_0^\infty \int_0^\infty g(\varepsilon)\frac{f(x-\varepsilon)}{\exp(\frac{x-\mu}{kT})-1}dxd\varepsilon, \quad (10)$$

$$N_0 = \frac{1}{\exp(-\frac{\mu}{kT})-1}, \quad (11)$$

The chemical potential can be found from (9) and (10) in the case without fluctuation as well as with fluctuations. The corresponding particle distributions over energy for low temperatures with taking into account condensate fraction are shown in figures (3) and (4) (compare with (1) and (2)).

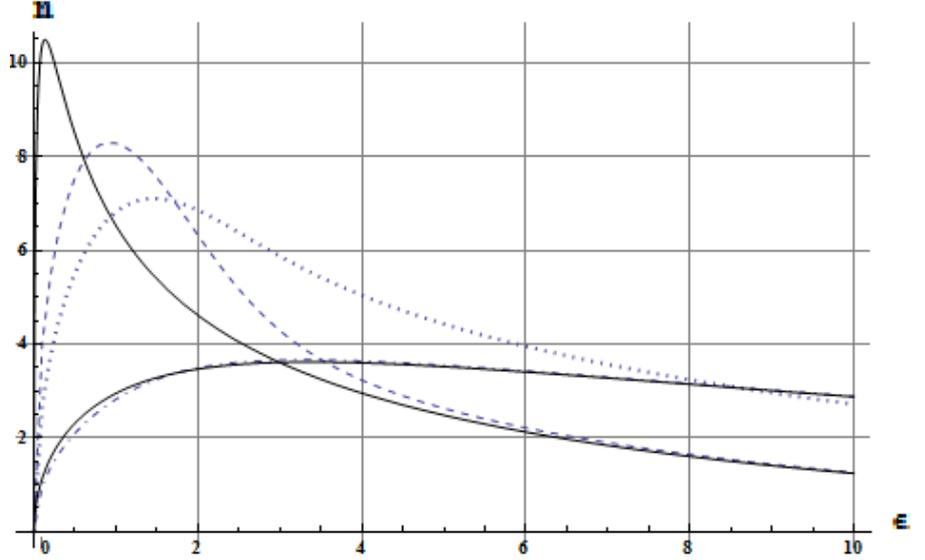

**Figure 3.** The particle number distributions over energy for different temperatures $T$. The upper solid line is the usual Bose-Einstein distribution at $T=8\varepsilon_0$, lower solid line is the usual Bose-Einstein distribution at $T=20\varepsilon_0$; dashed, dotted and chain lines show the distributions modified by fluctuations at temperatures $8\varepsilon_0, 14\varepsilon_0, 20\varepsilon_0$ respectively ($N=100$, $\sigma=1$).

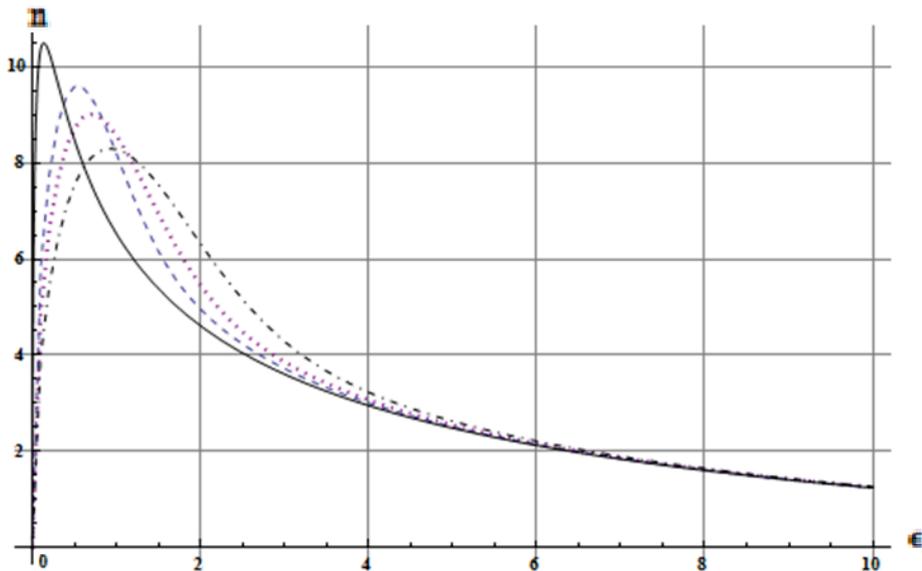

**Figure 4.** The modified by fluctuations particle distributions over energy for different values $\sigma$. The solid, dotted and chain lines of Bose-Einstein distributions at $\sigma = 0.5, 0.7, 1.0$, respectively ($N = 100$ and $T = 8\varepsilon_0$). For comparison the black color line shows the distribution of particles without fluctuations.

## 4. The temperature of Bose- Einstein condensation

The criterion for Bose-Einstein condensation is the equality of chemical potential to zero. When one writes convolution of Bose-Einstein distribution and Gaussian function the logarithmic divergence appears at chemical potential equal to zero. It suggests that for the considering the fluctuation effects we cannot take the chemical potential equal to zero, as it was done for an equilibrium system. For an ideal bose-gas consisting of $N$ particles with mass $M$ confined in a rigid container of volume $V$, condensation occurs at the temperature [13]:

$$T_C = \frac{h^2}{2\pi kM} \left[\frac{1}{2.612}\frac{N}{V}\right]^{2/3} \tag{12}$$

In the case of fluctuations the total number of particles is written as:

$$N = N_0(\theta) + \int_0^\infty \int_0^\infty \frac{g(\varepsilon)f(\varepsilon-x)}{Exp(\frac{x-\mu}{\theta})-1} d\varepsilon dx$$

$$= N_0(\theta) + \int_0^\infty dx\, n(x) \int_0^\infty d\varepsilon\, g(\varepsilon) f(x-\varepsilon) = N_0(\theta) + \int_0^\infty dx\, n(x) \widetilde{g}(x) \quad, \tag{13}$$

$$\int_0^\infty d\varepsilon\, g(\varepsilon) f(x-\varepsilon) = \widetilde{g}(x) \propto \sigma^{1/2}\Gamma(3/4)x^0 + \sigma^{-1/2}\Gamma(5/4)x^1 \tag{14}$$

$$N \propto N_0(\theta) + \int_0^\infty \frac{\sigma^{1/2}\Gamma(3/4)}{\exp(\frac{x-\mu}{\theta})-1} dx \propto \sigma^{1/2}\Gamma(3/4)\theta \int_0^\infty x^{-1} dx \to \ln x \Big|_0^\infty \tag{15}$$

Function $\widetilde{g}(x)$ has the linear dependence like $(ax+b)$, where $a$ and $b$ are the constants. The exponent in Bose-Einstein distribution can be expanded in Taylor series and the first non-vanishing term is proportional to $x/\theta$. Thus, we obtained the logarithmic divergence at $x=0$. If we set $\mu<0$ then the integral will be taken only numerically.

Let's consider the graph method of finding the condensation temperature. We are interested in the dependence of temperature Bose-Einstein condensation on fluctuation parameter [14]. Let's consider the grand canonical ensemble. We separate a very large term, corresponding to number of particles in the ground state:

$$N = N_0 + ZI(\mu), \tag{16}$$

where

$$Z = (\theta)^{3/2} \tag{17}$$

$$I(\mu) = \int_0^\infty \int_0^\infty \sqrt{\varepsilon}\, \frac{f(x-\varepsilon)}{\exp(x-\frac{\mu}{\theta})-1} dx d\varepsilon \tag{18}$$

All parameters in function $I(\mu)$ are measured in $\varepsilon_0$ units. Further, the equation (16) is divided into two parts. These parts contain the functions dependent on the temperature. This equation (the critical condition of finding the condensation temperature) can be solved graphically:

$$Z(\theta) = \frac{N}{I(\mu,\theta)} \tag{19}$$

The chemical potential in the fluctuation case is not equal to zero because of the divergence of the double integral (7). For each value of the chemical potential from the defined range we can numerically solve equation (19). Thus, we obtained the dependence of the chemical potential on the temperature. Extrapolating this curve to its intersection with the ordinate axis we got the value of the condensation temperature:

$$T_C = \frac{\varepsilon_0 * \theta}{k_b} \tag{20}$$

for the given fluctuation parameter at the point where the chemical potential is equal to zero.

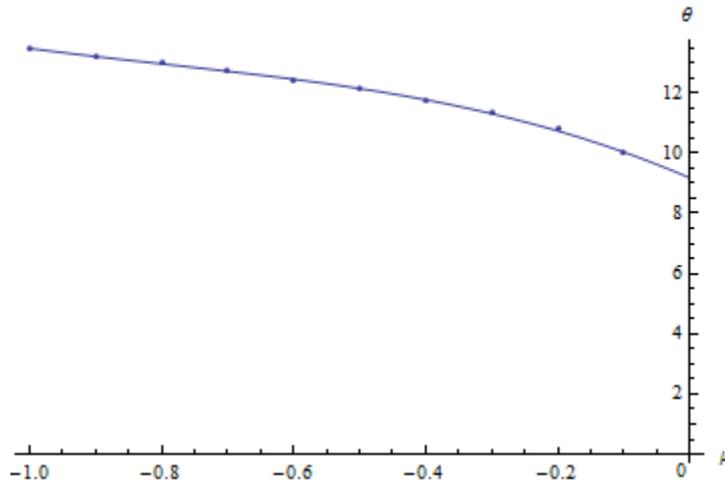

**Figure 5.** The extrapolated temperature dependence of the chemical potential with taking into account the fluctuations ($\sigma = 1$).

Thus, changing the fluctuation parameter, we can study the condensation temperature changes. The corresponding plot is shown in figure 6. If we extrapolate the curve of condensation temperature dependence on the fluctuation parameter, we shall obtain the value of the condensation temperature in point, where the fluctuation parameter is equal to zero. Following this procedure for the case of $^4$He atoms we have obtained the condensation temperature equal 3.12 K which is close to the theoretically calculated (3.07 K) [15].

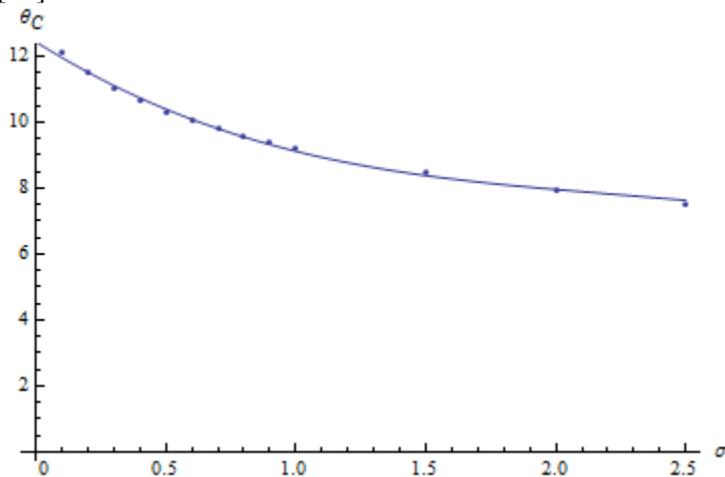

**Figure 6.** The condensation temperature versus the fluctuation parameter.

A more accurate method for the BEC temperature determination is the finding chemical potential from the condition (in dimensionless parameters normalized to $\varepsilon_o$:

$$N = \frac{1}{e^{-\bar{\mu}} - 1} + \tilde{\theta}^{3/2} \int_0^\infty \int_0^\infty \sqrt{\bar{\varepsilon}} f(\bar{x} - \bar{\varepsilon}) d\bar{\varepsilon} \frac{1}{e^{\bar{x} - \bar{\mu}} - 1} d\bar{x},$$
(21)

In the figure 7 the same data as in figure 6 are shown but in K degrees. Three points obtained by extrapolating the dependence of the chemical potential on the temperature (the value of chemical potential at different temperatures obtained from expression (21)) have been added for σ= 0.1, 1 and 2.5.

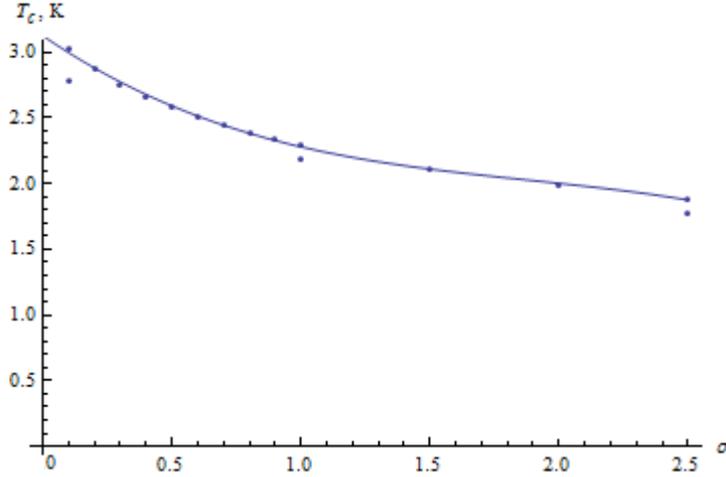

**Figure 7.** The solid line shows the dependence of the condensation temperature on the fluctuation parameter $\sigma$ (the dependence of the chemical potential on the temperature obtained by equation (19)). Three points show the dependence of the condensation temperature on the fluctuation parameter $\sigma$ = 0.1, 1 and 2.5 (the dependence of the chemical potential on the temperature obtained from (21)).

It can be concluded that the temperature of Bose-Einstein condensation decreases as the fluctuation parameter increases. The figure 7 shows that taking into account the fluctuations can give the value of the calculated condensation temperature more close to the experimental observed one (the experimental value 2.17 K for $^4$He can be obtained at the fluctuation parameter σ approximately equal to 2.4).

**Conclusion**
We have investigated in detail the effect of fluctuations in the form of Gaussian noise on the different physical quantities of an ideal bose-gas such as the chemical potential and the condensation temperature. The proposed simple model based on the convolution of Gaussian noise and Bose-Einstein distribution allowed to obtain particle distributions over energy for different temperatures and fluctuation parameter values. Using different methods, we have found the condensation temperature as the function of the fluctuations parameters.

**Acknowledgement**
The work was supported by the Ministry of Science and Education of Russia. A.P. thanks Region Pays de la Loire for partial individual financial support (MOA0020).

**References**
[1]     Yukalov V I  2004 Principal problems in Bose-Einstein condensation of dilute gases *Laser Physics Letters* **1** 435-461
[2]     Ziff R M ,Uhlenbeck G E and Kac M 1997, *Phys. Rev.* **32**, 169

[3]     Hauge E H 1969 *Physica Norvegica* **4**, 19
[4]     Fujiwara I, ter Haar D and Wergeland H 1970 *J. Stat. Phys*.**2**, 329
[5]     Politzer H D 1996 *Phys. Rev. A* **54**, 5048
[6]     Gajda M and Rzazewski K 1997  *Phys. Rev. Lett.* **78**, 2686; Navez P, Bitouk D, Gajda M, Idziaszek Z and  Rzazewski K 1997 *Phys. Rev. Lett.* **79**, 1789
[7]     Grossmann S and Holthaus M, Report No. *Cond.mat*/ 9709045 and Report No. *Cond.mat*/ 9709065
[8]     Giorgini S, Pitaevskii L P and Stringari S  1997 Anomalous fuctuations of the condensate in interacting Bose gases *Phys.Rev.Lett*. **80** 5040
[9]     Landau L D and Lifshitz E M 1980 *Statistical Physics* (United Kingdom:Pergamon/ Oxford)
[10]    Kleinert H 2000 *Phys. Rev. Lett.* **84**, 286
[11]    Landau L D and Lifshitz E M 2005 *Statistical physics* vol 5 (Moscow: Fizmatlit, 5th Edition) ed L P Pitaevskii p 616
[12]    Dorfman K E, Kim M and Svidzinsky A A 2011 Condensate statistics and thermodynamics of weakly interacting Bose gas: Recursion relation approach *Phys.Rev.A* **83** 033609
[14]    Bagnato V, Pritchard D E, Kleppner D 1987 Bose-Einstein condensation in an internal potential *Phys. Rev. A* **35** N 10 4354-58
[15]    Wang, Yongli Ma and Jizhou He 2011 Thermodynamics of finite Bose Systems: an exact canonical-ensemble treatment with different traps *J. Low Temp. Phys*.**162** 23–33
[16]    Wang Q A , Pezeril M, Le Mehaute A 2000  Nonextensive boson gas and specific heat of $^4$He superfuid *Physica A*  **278**  337-346
[17]    Lawani A A, Le Meur J, Tayurskii D, Kaabouchi A El, Nivanen L, Minisini B, Tsobnang F, Pezeril M, Le Mehaute A, Wang Q A 2007 Nonextensive approach to Bose-Einstein condensation of trapped interacting boson gas *J. of  Low Temperature Physics* **150**.N3-4 605-611